# Anti-Parity-Time Symmetric Optics via Flying Atoms

Peng Peng[1], Wanxia Cao[1], Ce Shen[1], Weizhi Qu[1], Jianming Wen[2]*, Liang Jiang[2]*, and Yanhong Xiao[1]*

[1]Department of Physics, State Key Laboratory of Surface Physics, and Key Laboratory of Micro and Nano Photonic Structures (Ministry of Education), Fudan University, Shanghai 200433, China

[2]Department of Applied Physics, Yale University, New Haven, Connecticut 06511, USA

*emails: jianming.wen@yale.edu; liang.jiang@yale.edu; yxiao@fudan.edu.cn.

**Abstract:** The recently-developed notion of 'parity-time (PT) symmetry' in optical systems with a controlled gain-loss interplay has spawned an intriguing way of achieving optical behaviors that are presently unattainable with standard arrangements. In most experimental studies so far, however, the implementations rely highly on the advances of nanotechnologies and sophisticated fabrication techniques to synthesize solid-state materials. Here, we report the first experimental demonstration of optical anti-PT symmetry, a counterpart of conventional PT symmetry, in a warm atomic-vapor cell. By exploiting rapid coherence transport via flying atoms, our scheme illustrates essential features of anti-PT symmetry with an unprecedented precision on phase-transition threshold, and substantially reduces experimental complexity and cost. This result represents a significant advance in non-Hermitian optics by bridging a firm connection with the field of atomic, molecular and optical physics, where novel phenomena and applications in quantum and nonlinear optics aided by (anti-)PT symmetry could be anticipated.

Canonical quantum mechanics postulates Hermitian Hamiltonians to describe closed physical systems so that the system energy is conserved with real eigenvalues and the orthogonality between eigenstates with different eigen-energy is ensured. For systems with open boundaries, non-Hermitian Hamiltonians with complex eigenvalues and non-orthogonal eigen-functions are commonly expected. However, a counterintuitive discovery by Bender and Boettcher in 1998[1] has radically challenged this cognition and evoked considerable efforts on extending canonical quantum theory to non-Hermitian Hamiltonian systems[2,3,4]. In their pioneering work[1], they showed that below some phase-transition point, a wide class of non-Hermitian Hamiltonians ($\hat{H}$) display entirely real spectra if they are invariant under the anti-linear parity-time (PT) operator ($\hat{P}\hat{T}$), $[\hat{H}, \hat{P}\hat{T}] = 0$. In non-relativistic quantum mechanics, governed by Schrödinger equation, a necessary but not sufficient condition[1] for PT symmetry to hold mandates the complex potential involved to satisfy $V(x) = V^*(-x)$, which implies its real (imaginary) part is an even (odd) function of position. Even more exciting is the occurrence of a sharp, symmetry-breaking transition once a non-Hermitian parameter crosses an exceptional point. As such, the Hamiltonian and PT operator no longer share the same set of eigen-functions and the real eigen-spectra of the system start to become complex.

Despite the ramifications of these theoretical developments[1,2,3,4], unfortunately, quantum mechanics is by nature an Hermitian theory and thus any attempt to observe PT symmetry in such systems is out of reach. On the other hand, due to the presence of gain and loss, optics has been recognized as a fertile ground for experimental investigations of PT symmetry. Given such optical settings, they can be fully realized without introducing any conflict with the Hermiticity of standard quantum mechanics. Thanks to the mathematical isomorphism existing between the quantum Schrödinger and the paraxial wave equations, a PT-symmetric complex potential can be readily established by judiciously engaging refractive indices with balanced gain and loss[5], $n^*(x) = n(-x)$. Optical realizations[6,7,8,9,10,11,12,13,14] of PT symmetry have motivated various synthetic designs exhibiting peculiar properties that are otherwise unattainable in traditional Hermitian structures. Among them, is the possibility for band merging effects[5], double refraction[15], nonreciprocal light propagation[7,10,16,17,18], and power oscillations[7,19]. Some useful schemes have also been put forward to coherent perfect laser-absorber[11,20,21], optical switches[22], optical couplers[23], and single-mode amplifiers[14,24]. Enlightened by optical settings, other systems using plasmonics[25], LRC circuits[26], acoustics[27], artificial lattices[28] and optomechanics[29] have also been reported successively.

As a counterpart of standard PT symmetry, interestingly, an anti-PT symmetric Hamiltonian satisfies the anti-commutation relation with the $\hat{P}\hat{T}$ operator, $\{\hat{H}, \hat{P}\hat{T}\} = 0$. Mathematically, multiplying a conventional PT-symmetric Hamiltonian by "$i$" would render it anti-PT symmetric, indicating that anti-PT optical systems exhibit properties completely conjugate to those of PT systems. Alternatively, optical properties for the imaginary part of a PT-symmetric potential now emerge in the real part of its corresponding anti-PT-symmetric potential. For instance, before phase breaking lossless propagation in a PT system would correspond to dispersiveless propagation in an anti-PT system. Such intriguing effects may open up new opportunities for manipulating the refractive indices of light, and form a necessary complementary probe in non-Hermitian optics. In spite of these appealing features, its realization is experimentally challenging. We notice that the first theoretical proposal[30] on anti-PT symmetry considers a composite system of metamaterials by demanding $n^*(-x) = -n(x)$, which however compromises an impractical balance of positive and negative real refractive indices. Its replication is hereafter brought forward with optical lattices of spatially driven cold atoms[31]. In fact, theoretical efforts[32,33,34] on realizing PT symmetry in cold atoms have already been put forward in the spatial domain. Coherently-prepared multi-level atoms are indeed attractive systems for exploiting (anti-)PT-symmetric optics, because of their easy reconfiguration, flexible tunability, and especially various coherence control techniques enabled by electromagnetically induced transparency (EIT)[35,36,37,38,39,40]. Yet, to date, no experiment based upon those proposals[31,32,33,34] has been realized, mainly due to the difficulty of creating PT or anti-PT symmetric optical potentials through rather complicated spatial modulations of driving fields or optical lattices.

By introducing a distinct coupling mechanism, in this paper we report the first experimental realization of anti-PT symmetric optics. In contrast to previous proposals, here an anti-PT symmetric non-Hermitian Hamiltonian is effectively constructed via mediating diffused atomic coherence carried by hot moving atoms. The precise measurement on (anti-)PT phase transition in the frequency domain further differentiates our work from the formers rooted on either solid-material or (theoretical) cold-atomic systems. The high resolution on anti-PT threshold achieved here would be very challenging for solid material schemes. Besides its superior performance, the characteristics of simple operation, versatile tunability, easy controllability, and low cost render the scheme a perfect stage for probing non-Hermitian optics at the exceptional point[41,42,43,44]. Most importantly, all experimental explorations on PT-symmetric optics so far depend exceedingly upon gain-loss arrangements with use of manmade solid-state materials and rely heavily on sophisticated nanofabrication technologies, which are very demanding even in state-of-the-art laboratories. Therefore, by harvesting hot atoms our proof-of-principle illustration on anti-PT symmetry without gain opens a new avenue for scrutinizing PT symmetry and bridges a firm connection with the field of atomic, molecular and optical (AMO) physics.

Our experiments (schematically shown in Fig. 1) are accomplished through fast coherence transport in a wall-coated warm $^{87}$Rb vapor cell (2.5 cm in diameter and 5 cm in length) without buffer gas. The inner surface of the cell is coated by coherence-preserving paraffin[45] which allows atoms to undergo thousands of wall collisions with little demolition of their internal quantum state. The cell is set at the temperature of 40°C to maintain an optically thin medium, and housed within a four-layer magnetic shield to screen out ambient magnetic field. Inside the shield a solenoid is used to generate a uniform magnetic field for the sweep of the EIT spectra. An external cavity diode laser operating at the $^{87}$Rb D$_1$ line (795 nm) provides the light for the probe and control fields. The laser beam is spatially split into four beams (1.2 mm in diameter) using half-wave plates and polarization beam splitters (PBS). Orthogonally polarized probe and control light are recombined and directed into two optical channels (Ch1 and Ch2) separated transversely by 1 cm. The right circularly-polarized strong control fields (of the power ~180 µW) are on resonance with the transition $|1\rangle \to |3\rangle$, and the two left circularly-polarized weak probe (of the power ~3 µW) are almost resonant with $|2\rangle \to |3\rangle$ but frequency shifted oppositely by the same one-photon detuning $|\Delta_0|$ using acousto-optical modulators (AOM). In the three-level $\Lambda$-type structure of interest, the co-propagating probe and control fields set up the standard EIT effect and establish a long-lived ground-state coherence or a collective spin wave in each channel. The two spin waves are effectively coupled through the ballistic motion of $^{87}$Rb atoms in the cell. The randomness and irreversibility lays the foundation of interrogating the non-Hermiticity of the effective Hamiltonian for the evolution of the probe dynamics through two indirectly-coupled channels. EIT spectral lineshapes are measured by gradually sweeping the magnetic field while recording the transmission of the output continuous wave probe light.

The atom-light interaction in the current isotropic system is computed by the density matrix formulism[46] (Supplementary Information). Under certain approximations, the coupled equations between the two ground-state coherence in Ch1 and Ch2 effectively give rise to the following non-Hermitian Hamiltonian

$$H = \begin{bmatrix} -|\Delta_0| - i\gamma'_{12} & i\Gamma_c e^{2i|\Delta_0|t} \\ i\Gamma_c e^{-2i|\Delta_0|t} & |\Delta_0| - i\gamma'_{12} \end{bmatrix}. \tag{1}$$

Here, $\gamma'_{12} = \gamma_{12} + \Gamma_c + 2\Gamma_p$, with $\gamma_{12}$ the dephasing rate of the ground-state coherence, $\Gamma_c$ the coherence coupling rate between the two channels, and $2\Gamma_p = 2|\Omega_1|^2/\gamma_{31}$ the total pumping rate by the two control beams with the same Rabi frequency $\Omega_1$, where $\gamma_{31}$ is the atomic optical coherence decay rate.

For simplicity of analysis but without loss of generality, we consider the effective Hamiltonian (1) at periodically-distributed discrete time points satisfying $e^{2i|\Delta_0|t} = 1$, which allows eqn. (1) to be reduced to the form:

$$H_{eff} = \begin{bmatrix} -|\Delta_0| - i\gamma'_{12} & i\Gamma_c \\ i\Gamma_c & |\Delta_0| - i\gamma'_{12} \end{bmatrix}, \tag{2}$$

associated with two eigen-EIT supermodes

$$\omega_\pm = -i\gamma'_{12} \pm \sqrt{\Delta_0^2 - \Gamma_c^2}. \tag{3}$$

As the essential results of this work, eqns. (1)-(3) contain rich physics and possess many interesting properties. First of all, the above Hamiltonian (1) complies with $\hat{P}\hat{T}H = -H$, in contrast to $\hat{P}\hat{T}H = H$ for conventional PT symmetry. To emphasize such a counterpart to common PT symmetry, we term the Hamiltonian ascribed above anti-PT symmetric. Furthermore, such a Hamiltonian leads to an intriguing phase transition reflected on the two eigen-EIT spectral branches. Specifically, in the regime of the unbroken anti-PT phase ($|\Delta_0| < \Gamma_c$), the two EIT resonances coincide at the center $\omega = 0$ but with different linewidths. The spontaneous anti-PT symmetry breaking point occurs at the critical point $|\Delta_0| = \Gamma_c$ where the two supermodes are perfectly overlapped. When $|\Delta_0| > \Gamma_c$, the system is in the symmetry-broken regime, and the line-centers of the two eigen-spectra obey a quadratic curvature, resembling a passively-coupled system.

The spectral profiles of the two eigen-EIT super modes can be probed in our experiment by sweeping the magnetic field to measure the weak-probe transmission spectra for constant-intensity probe and control inputs. With all lasers on, indeed, the EIT spectra (Fig. 2) display a beating signal oscillating at frequency $|2\Delta_0|$ as expected from eqn. (1). The physics origin of such a beating behavior is the following. Due to the presence of the oppositely detuned probes in Ch1 and Ch2, the locally-created spin waves within the laser-beam volumes oscillate with $e^{i|\Delta_0|t}$ and $e^{-i|\Delta_0|t}$, respectively. Meanwhile, during the formation process of the steady state coherence in each channel, atomic motion at a much faster time scale redistribute and mix these two spin waves within the entire vapor cell. At steady state, the spin wave in each channel thus contains both oscillating components and yields oscillatory patterns with a beating frequency $|2\Delta_0|$ manifested in the probe transmission spectrum for each fixed $\delta_B$ (Supplementary Information). In order to observe these beating signals without being washed out during time averaging, in the experiment we ensure that the magnetic field is swept slowly enough and its sweeping periodicity is maintained at integer times of the beating period. Figure 2 shows two sets of representative probe transmission spectra with $\Delta_0 = \pm 15$ Hz (Fig. 2a) in the unbroken phase regime and $\Delta_0 = \pm 60$ Hz (Fig. 2b ) in the broken phase regime. It is worthwhile to emphasize that the horizontal axis here is a function of both $\delta_B$ and time. The high contrast of the observed beating patterns is attributed to motional averaging in our system (see Supplementary Information). In order to quantitatively compare the extracted eigen-EIT spectra with the theory, in each subplot, for convenience, we choose a phase reference point, which is the peak of the beat note with the maximal beating amplitude (highlighted in dark blue) and satisfies $e^{i(2|\Delta_0|t+\Delta\varphi)} = 1$, where $\Delta\varphi$ is the difference between the two channel's probe-control relative phases (see Supplementary Information). Starting from this reference point, the rest discrete time points (red dots) are sequentially identified at distances of integral multiples of the beating period. The extracted EIT spectra formed by these red dots are a linear superposition of the two eigen-EIT supermodes as given in eqn.(3). As we can see, good agreement is achieved between theory and experiment. For $\Delta_0 = \pm 15$ Hz the two eigen-EIT supermodes are coalesced at $\delta_B = 0$, while for $\Delta_0 = \pm 60$ Hz, the two modes are pulled closer to each other without complete coincidence.

The evolution of the anti-PT supermodes is carefully examined by varying $|\Delta_0|$. By applying the method elaborated above, in Fig. 3 the extracted line centers and their corresponding linewidths are plotted as a function of $|2\Delta_0|$, showing an excellent agreement with theoretical predictions. Remarkably, the current system exposes an extraordinary resolution on phase transition down to the Hertz level, which would be highly demanding for solid-state materials. The experimental data in Fig. 3 evidently reveals the phase-breaking point at $|2\Delta_0| = 30.5$ Hz, implying the coherence coupling rate $\Gamma_c$ of about 15 Hz. According to eqn. (3), when $\Delta_0 = 0$ the full-linewidth difference between the two supermodes is equal to $4\Gamma_c$, which is verified by the fitting curve to the experiment data (Fig. 3a). On the other hand, eqn. (3) predicts that the full linewidth of the lower eigenmode near $\Delta_0 = 0$ should be $2(2\Gamma_p + \gamma_{12})$. To experimentally attest this prediction, we first allow only one channel on and obtain $\gamma_{12} = 5$ Hz through the zero-power EIT linewidth measurement. By virtue of the recorded single-channel EIT (not shown) full-width of $2(\Gamma_p + \gamma_{12}) = 69$ Hz we deduce $2\Gamma_p = 59$ Hz. Now $2(2\Gamma_p + \gamma_{12})$ should read 128 Hz, concurring with the width of the lower eigenmode for $\Delta_0 = 0$ as read from Fig. 3a.

The anti-PT system demonstrated here provides a new and simple way for nonlinear optics at low-light level. Due to the small nonlinearities in crystals, there has been great interests in pursuing low-light-level nonlinear optics using atoms. Existing approaches usually involve nonlinear properties inherent to a multi-level atomic structure configuration. Interestingly, our anti-PT symmetric system provides another possibility to establish effective interaction between two weak light beams via atomic coherence transport. As witnessed above in the phase-unbreaking regime, the two output probes' EIT in Ch1 and Ch2 converge regardless of their originally separated spectra. Alternatively, the existence of the weak probe beam in one channel significantly influences the transmission (and the refractive index) of the weak probe in the other channel. Such an effect becomes most dramatic near the critical point. By utilizing this effect, Fig. 4 reports our proof-of-principle demonstration based upon anti-PT symmetry at the low-light level. The two weak probe fields are frequency detuned with respective to the control fields to opposite directions by 14 Hz. In this measurement, we have averaged out the beating in the EIT spectra by slightly offsetting the sweeping period of the magnetic field from integer multiples of the beating period. The blue curve stands for the EIT spectrum of the Ch1 output probe as the Ch2's control on but probe off. The red curve represents the output spectrum of the Ch1 probe by turning the Ch2's probe on, where a notable shift of the resonance peak is clearly observed. Despite that in the current experiment the input probe power cannot be lower than 50 nW due to technical noises, in principle, such an idea may be feasible even at the single-photon level, which would extend the potential applications of anti-PT symmetry towards single-photon nonlinear optics and the implementation of a single-photon phase gate.

In conclusion, we have reported a novel platform for studying non-Hermitian optics in a warm atomic vapor cell by taking advantage of rapid coherence transport. As an example, we utilized such a system to generically realize anti-PT symmetric optics for the first time, and achieved unprecedented accuracy on phase-transition measurements. The gainless scheme considered here provides an ideal tool for the investigation of super- and sub-luminal light propagation. The structureless configuration also makes the system perfect for studying transverse optical properties of probe beams above and below anti-PT phase breaking. Further discussion on the Kramers-Krönig relations[47] along with these topics will be addressed elsewhere. It is also possible to form (anti-)PT symmetry in the spatial domain. Given the tight connection of our system with slow/fast light and quantum memory[47], we anticipate that new directions could be opened up after this work. In particular, it may become a fruitful arena, by bridging (anti-)PT symmetry with the AMO field, for other research areas such as quantum optics and quantum information science. Despite that the mediator analyzed in this work is freely moving atoms, similar strategies might be extended to other systems where the indirect coupling is established through electrons, plasma, or phonons.

**Acknowledgements**

This research was supported by the NBRPC (973 Program Grants No. 2012CB921604 and No. 2011CB921604), NNSFC (Grant No. 11322436), and the Research Fund for the Doctoral Program of Higher Education of China. J.W. and L.J. acknowledge funding support from the ARO, the AFSOR MURI, the DARPA QUINESS program, the Alfred P. Sloan Foundation, and the David and Lucile Packard Foundation.

List of Figures

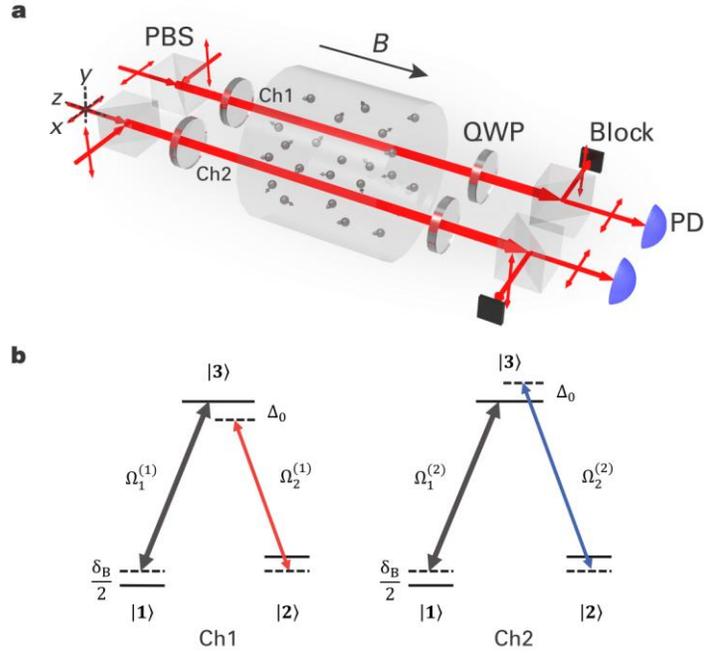

**Figure 1 | Anti-PT symmetric optics via rapid atomic coherence transport in a warm $^{87}$Rb vapor cell. a**, Schematic of 3D view of the system. In two spatially separated optical channels (Ch1 and Ch2), each contains collinearly propagating weak probe and strong control fields operating under the condition of EIT. Ballistic atomic motion in a paraffin wall-coated vacuum cell distributes atomic coherence through the optically thin medium and results in effective coupling between two optical channels. With the temperature retained at 40°C, the cell is housed within a four-layer magnetic shield to screen out external magnetic field. Output probe transmission spectra are measured by sweeping a homogeneous magnetic field generated by a solenoid inside the shield. **b**, The implementation of anti-PT symmetry utilizing standard three-level Λ-type EIT configurations in two channels. An external cavity diode laser tuned at the $^{87}$Rb D$_1$ line provides the light for the probe and control fields with orthogonal circular polarizations. Right circularly-polarized strong control fields, $\Omega_1^{(1)}$ and $\Omega_1^{(2)}$ in Ch1 and Ch2 are, respectively, on resonance with the transition $|1\rangle \to |3\rangle$ while left circularly-polarized weak probe fields, $\Omega_2^{(1)}$ and $\Omega_2^{(2)}$ in Ch1 and Ch2, are near resonant with $|2\rangle \to |3\rangle$ but frequency shifted oppositely by a same amount $|\Delta_0|$, with $\Delta_0$ the probe field frequency minus the control field frequency. At steady state, all the population is mainly on the ground state $|2\rangle$.

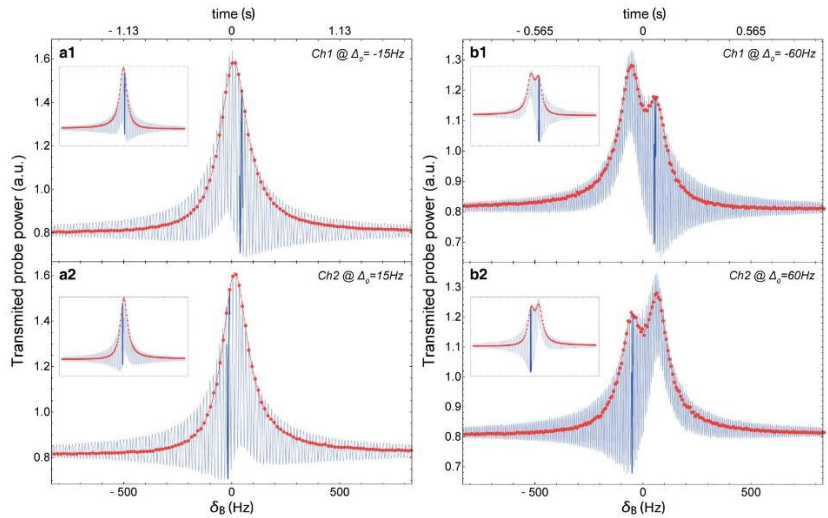

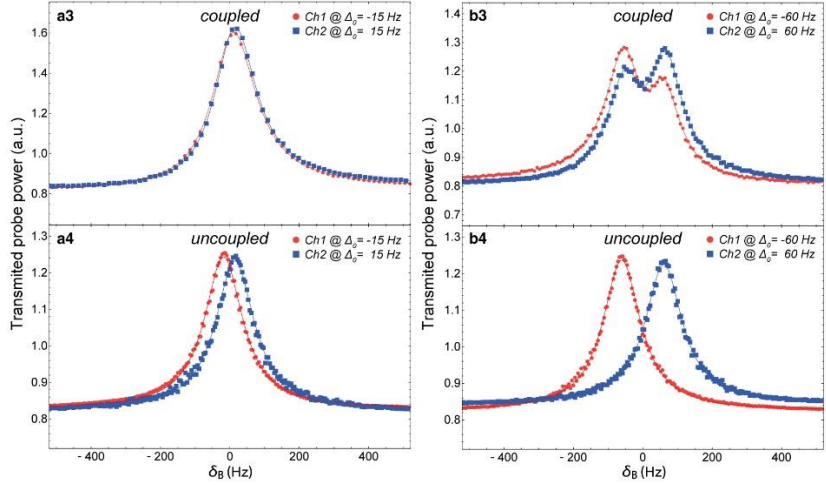

**Figure 2 | Representative transmission spectra of output probe light in anti-PT symmetry. a1 and a2,** In the regime of unbroken anti-PT symmetry, typical transmission spectra of output probe light after Ch1 (with the probe red detuned from the control by ~15 Hz) and Ch2 (with the probe blue detuned from the control by ~15 Hz) exhibit beating frequency ~30 Hz. **b1 and b2,** In the regime of broken anti-PT symmetry, typical probe transmission spectra with probe detuning ±60 Hz in Ch1 and Ch2, respectively. The bold, dotted curves are two anti-PT eigen-EIT modes extracted at certain time points $t$ satisfying $e^{i(2|\Delta_0|t+\Delta\varphi)} = 1$, and the marked dark-blue beating notes in the upper four figures represent the time reference point (Supplementary Information). The insets are theoretical simulations. **a3 and b3,** The dotted curves in Ch1 (**a1** and **a2**) and Ch2 (**b1 and b2**) are plotted together for comparison. **a4 and b4,** In contrast, uncoupled EIT spectra separately measured from Ch1 (with red detuning ~15 Hz/~60 Hz) with both channels' control and only Ch1 probe input on, and from Ch2 (with blue detuning ~15 Hz/~60 Hz) with both control and only Ch2 probe input on. The parameters here are: control powers of ~180 μW and probe powers of ~3.7 μW.

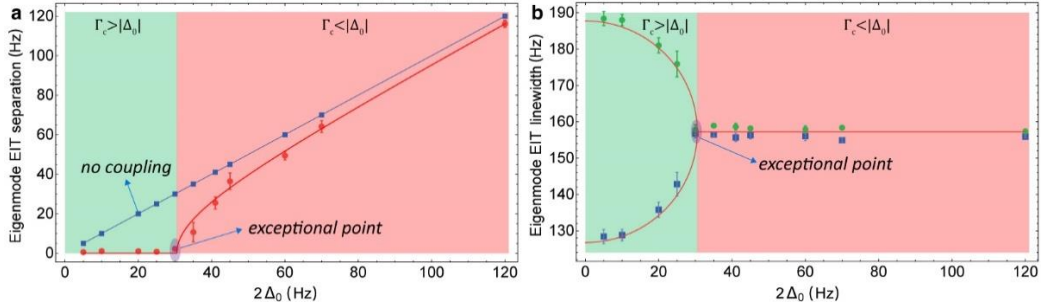

**Figure 3 | Anti-PT symmetry breaking in coupled EIT channels in a homogeneous, warm $^{87}$Rb vapor cell. a and b,** The real part Re($\omega$) and imaginary part Im($\omega$) of the two eigen-frequencies of coupled-EIT supermodes as a function of the probe detuning $|\Delta_0|$. The data points are obtained from curve-fitting the measured transmission spectra to the theoretical result (3). In **a**, as a comparison, the blue solid squares represent the EIT peak separation between two uncoupled channels, and the red round dots are for the case of both Ch1 and Ch2 on. In **b**, the green round dots and blue solid squares are the extracted linewidths of the two eigen EIT modes, respectively. As described in the text, the linewidth values here are in an excellent agreement with an independent check. The experiment parameters here are the same as those in Fig. 2.

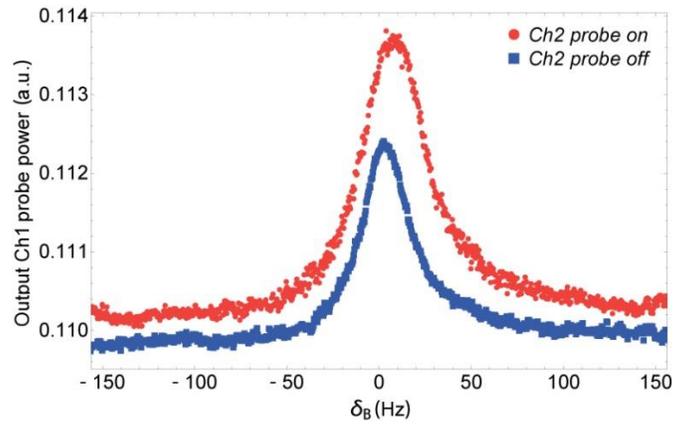

**Figure 4 | Demonstration of effective interaction between two weak probe fields aided by anti-PT symmetry in the unbroken-phase regime.** In the presence of both control beams in the two channels, the EIT spectra of the output probe beam from Ch2 with the Ch1 probe off (blue curve) and on (red curve) are obtained at the low-light level for the probes. It is evident that the presence of the Ch2 probe clearly shifts the resonance position of the Ch1 probe. The parameters are: control powers of ~30 μW and probe powers of ~50 nW.